\documentclass[prb,onecolumn]{revtex4}
\usepackage{eurosym}
\usepackage{amsfonts}
\usepackage{amsmath}
\usepackage{amsopn}
\usepackage{amssymb}
\usepackage{graphicx}
\usepackage{graphics}
\usepackage{color}
\usepackage{verbatim}

\begin{document}

\title{Two-dimensional symbiotic solitons and vortices in binary condensates
with attractive cross-species interaction}
\author{Xuekai Ma$^{1}$, Rodislav Driben$^{1}$, Boris A. Malomed$^{2}$,
Torsten Meier$^{1}$, and Stefan Schumacher$^{1}$}
\affiliation{$^{1}$Department of Physics and Center for Optoelectronics and Photonics
Paderborn (CeOPP), Universit\"{a}t Paderborn, Warburger Strasse 100, 33098
Paderborn, Germany}
\affiliation{$^{2}$Department of Physical Electronics, School of Electrical Engineering,
Faculty of Engineering, Tel Aviv University, Tel Aviv 69978, Israel.}

\begin{abstract}
We consider a two-dimensional (2D)\ two-component spinor system with cubic
attraction between the components and intra-species self-repulsion, which
may be realized in atomic Bose-Einstein condensates, as well as in a
quasi-equilibrium condensate of microcavity polaritons. Including a 2D
spatially periodic potential, which is necessary for the stabilization of
the system against the critical collapse, we use detailed numerical
calculations and an analytical variational approximation (VA) to predict the
existence and stability of several types of 2D \textit{symbiotic solitons}
in the spinor system. Stability ranges are found for symmetric and
asymmetric symbiotic fundamental solitons and vortices, including \textit{%
hidden-vorticity} (HV) modes, with opposite vorticities in the two
components. The VA produces exceptionally accurate predictions for the
fundamental solitons and vortices. The fundamental solitons, both symmetric
and asymmetric ones, are \emph{completely stable}, in either case when they
exist as gap solitons or regular ones. The symmetric and asymmetric vortices
are stable if the inter-component attraction is stronger than the
intra-species repulsion, while the HV modes have their stability region in
the opposite case.
\end{abstract}

\maketitle



\section*{Introduction}

Multidimensional localized modes in nonlinear dispersive media, which are
usually considered as solitons, is a topic of great interest in many areas
of physics \cite%
{general-reviews,Ostrovskaya-2012-pra,Sich-2012-npho,Amo-2011-Science,
Egorov-2009-prl,Lagoudakis-2008-nphy}, especially in nonlinear photonics
and in the studies of matter waves in Bose-Einstein condensates (BECs). In
addition to their significance to fundamental studies, two- and
three-dimensional (2D and 3D) solitons offer potential applications, such as
the use of spatiotemporal optical solitons as bit carriers in
data-processing schemes \cite{Wagner} and the design of matter-wave
interferometers using 3D matter-wave solitons \cite{interferometry}.

Unlike 1D solitons, which are usually stable objects \cite{Michel},
stability is a major issue for their 2D and 3D counterparts. The ubiquitous
cubic self-focusing nonlinearity, which readily produces formal 2D and 3D
soliton solutions, gives rise to the wave collapse (\textit{critical}\
\textit{collapse} in 2D and \textit{supercritical collapse} in 3D) \cite%
{collapse,VK}, which entirely destabilizes the respective soliton families.
For this reason, the first example of solitons which was predicted in
nonlinear optics, \textit{viz}., the \textit{Townes solitons} \cite{Townes},
i.e., 2D self-trapped modes supported by a cubic self-focusing nonlinearity, are
subject to the instability caused by the critical collapse, which is why they
have not been created in experiment. Multidimensional solitons with
intrinsic vorticity (vortex rings), which also formally exist in cubic media
(in particular, vortex counterparts of the Townes solitons \cite{Minsk}),
are subject to an even stronger splitting instability initiated by azimuthal
perturbations \cite{general-reviews}.

Thus, the stabilization of multidimensional fundamental and vortex solitons
is an issue of great importance in various physical settings. In a 2D
geometry with cubic self-focusing nonlinearity, the stabilizing factor must
break the specific scale invariance, which underlies the critical collapse.
The most straightforward possibility for that is provided by the addition of
a spatially periodic (lattice) potential, which fixes a particular spatial
scale (the lattice period). As a result, it was predicted that such
potentials provide for the stabilization of both fundamental and vortex
solitons \cite{BBB,BBB-hex}. In most cases, stable vortices are created in
the form of a ring-shaped chain of local density peaks, with the vorticity
carried by the superimposed phase profile which features a phase circulation
of $2\pi S$, with integer $S$ being the respective topological charge.

In many cases, the natural interactions in the underlying physical media are
repulsive, hence they cannot create regular solitons, which play the role of
the ground state in the physical system, although it is possible to create
gap solitons as a result of the interplay of the lattice potential and
a repulsive nonlinearity \cite{GS,gap-vortex,gap-vortex2}. In particular,
inter-atomic forces in bosonic gases are normally repulsive \cite{dummy}, as
well as the exciton-exciton interactions in polariton condensates in
semiconductor microcavities \cite{dummy2}.

The overall repulsive interaction may be switched into attraction by means
of a Feshbach resonance (FR), in bosonic gases \cite{Chin-2010-RMP} and
polariton condensates \cite{Takemura-2014-NP} alike. The FR applies to the
inter-component interactions in binary (spinor or pseudo-spinor) condensates
\cite{binary}. This suggests a new approach to the creation of solitons in
spinor condensates: while each component features self-repulsion, the
FR-induced attraction between them makes it possible to create \textit{%
symbiotic solitons}, supported solely by the attraction between the two
components, which may even overcome the intrinsic repulsion in each of them
\cite{symbio}. In particular, for microcavity polaritons this condition may
be satisfied in a narrow spectral range, as explained in more detail below.
The formation of localized symbiotic modes in the presence of a lattice
potential was studied too, but only in 1D settings \cite{semigap,Thawatchai}.

The objective of the present work is to develop the concept of symbiotic
solitons for 2D spinor condensates with an inter-component cubic attraction
and intra-species repulsion. To secure the stability of the 2D solitons
against the critical collapse, the system must include a lattice \
potential, which can be easily implemented in experiments for both atomic
BEC (as an optical lattice) \cite{dummy} and polariton condensates \cite{UK}
(in the latter work, the lattice was used for the experimental creation of
2D exciton-polariton solitons).

The results presented below are obtained by means of a combination of an
analytical variational approximation (VA) and a systematic numerical
investigation of the existence, stability, and dynamics of the solitons. We
find that the VA provides very good accuracy in predicting fundamental and
vortex solitons, both symmetric and asymmetric ones, with respect to the two
components. We also consider self-trapped modes with \textit{hidden vorticity%
} (HV), i.e., a soliton with opposite signs of the vorticity in its two
components \cite{Brtka-2010-pra}. The solution families include both gap
solitons and regular ones, as concerns their placing relative to the bandgap
spectrum of the linearized system. It is found that all the fundamental
solitons are \emph{completely stable} solutions, while the vortices have a
stability region when the cross-component attraction is stronger than the
intra-species repulsion. The HV modes may be stable in the opposite case.


\section*{Results}

\subsection*{The model}

The dynamics of both two-component atomic BEC and coherent microcavity
polariton gases (with losses appropriately compensated by gain), are
modeled by the symmetric pair of coupled nonlinear Schr\"{o}dinger
equations \cite{dummy,dummy2},
\begin{subequations}
\label{model}
\begin{align}
i\frac{\partial \phi }{\partial t}& =-\nabla ^{2}\phi +|\phi |^{2}\phi
-g|\psi |^{2}\phi +V\left( x,y\right) \phi \,,  \label{model:A} \\
i\frac{\partial \psi }{\partial t}& =-\nabla ^{2}\psi +|\psi |^{2}\psi
-g|\phi |^{2}\psi +V\left( x,y\right) \psi \,.  \label{model:B}
\end{align}%
Here $\phi \left( x,y,t\right) $ and $\psi \left( x,y,t\right) $ are the
coherent fields of the two components, with $\nabla ^{2}\equiv \partial
^{2}/\partial x^{2}+\partial ^{2}/\partial y^{2}$ being the 2D Laplacian and
$V(x,y)$ an effective lattice potential. It is assumed, as noted above, that
the self-interaction of each component is repulsive, while the
cross-interaction is attractive, accounted for by $g>0$; another mechanism
that may give rise to effective attraction in the polariton fluid was
proposed in Ref. \cite{Dominici-2015-NC}. The equations are written in the
scaled form, so that the coefficients in front of the Laplacian and
self-repulsion terms are normalized to unity. Equal coefficients in front of
the Laplacians in both equations imply that the spinor describes a pair of
different hyperfine states of the same atom in the bosonic gas. The same
symmetry refers to the two spin states of excitons in the polariton gas \cite%
{Lagoudakis-2009-Science}.

\begin{figure}[t]
\centerline{\includegraphics[width=14cm]{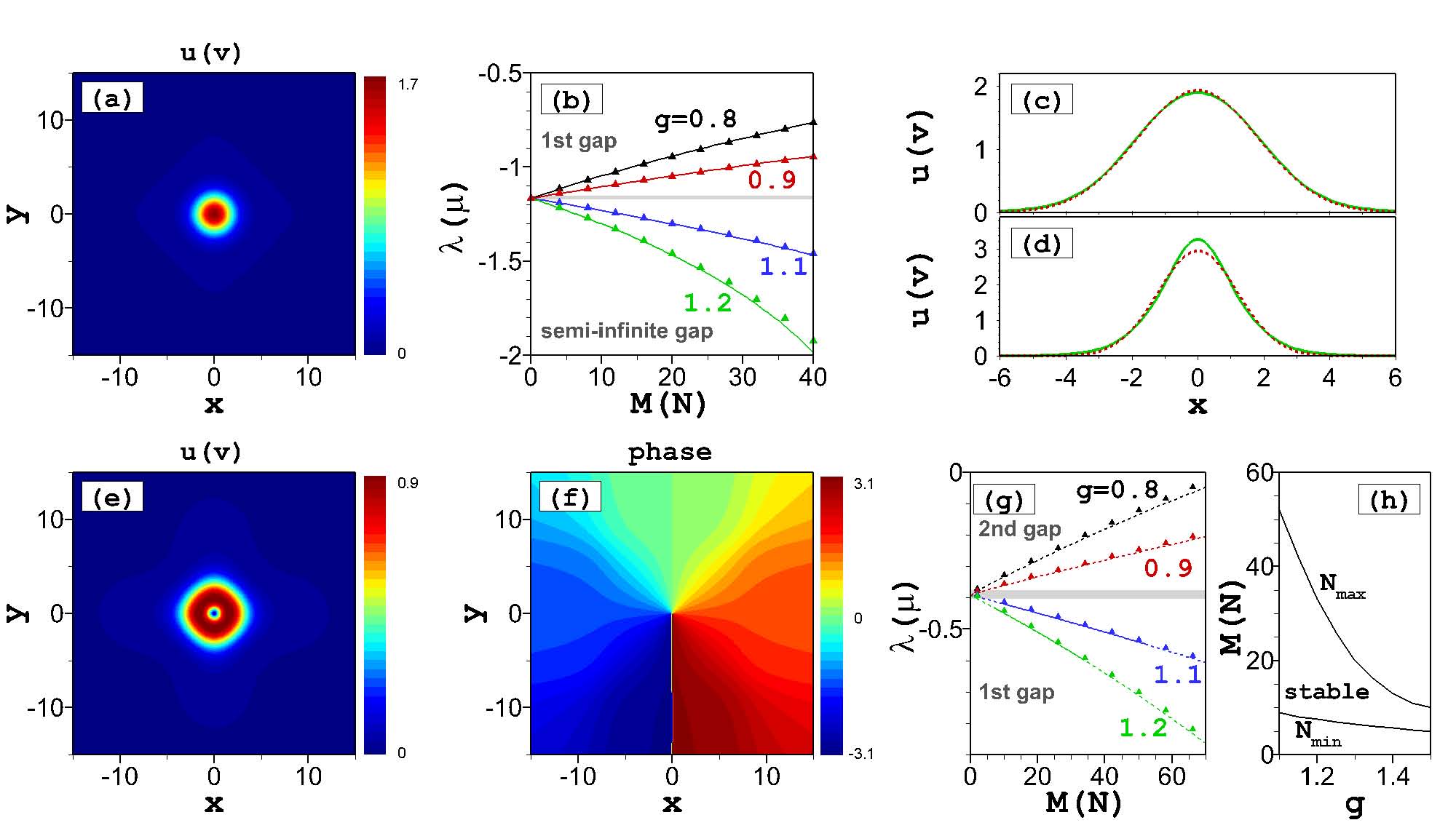}}
\caption{\textbf{Symmetric solitons and vortices.} (a) The amplitude
distribution in a stable symmetric fundamental soliton, for $g=1.1$ and $M=20
$. (b) Dependence of the norm ($M\equiv N$) on the chemical potential ($%
\protect\lambda \equiv \protect\mu $) for the family of fundamental
symmetric solitons at different values of the cross-attraction coefficient, $%
g$. Solid lines depict stable numerically found solutions, while triangles
represent the corresponding results produced by the VA as per Eqs. (\protect
\ref{ab}) and (\protect\ref{lm}). (c,d): Comparison of numerically found
(solid lines) and VA-predicted (dashed lines) profiles of the fundamental
symmetric solitons at $g=0.8$, $M=40$ (c) and $g=1.2,$ $M=40$ (d). Panels
(e) and (f) display, respectively the amplitude and phase distributions in a
numerically found stable symmetric vortex solitons for $g=1.1$ and $M=20$.
(g) Dependence of the norm ($M\equiv N$) on the chemical potential ($\protect%
\lambda \equiv $ $\protect\mu $) for the symmetric vortex solitons for
different values of $g$. Solid and dashed lines represent stable and
unstable vortices, respectively, while triangles depict the corresponding VA
results, as per Eqs. (\protect\ref{abV}) and (\protect\ref{lmV}). At $g>1$,
the vortices are stable in a finite interval of values of the norm, see Eq. (%
\protect\ref{NNN}). (h) Dependence of the stability
boundaries of vortices, $N_{\max }$ and $N_{\min }$, on $g$. The gray
regions in (b) and (g) represent the (narrow) first and second bands of the
lattice-potential spectrum, respectively.}\label{symmetric}
\end{figure}

The lattice potential is taken in the usual form, $V=-\left[ \cos \left(
2\pi x/P\right) +\cos \left( 2\pi y/P\right) \right] $, with spatial period $%
P$. Below, generic numerical results are adequately presented for $P=10$.
The depth of the potential, $V_{\max }-V_{\min }=4$, is fixed by means of
the remaining scaling invariance. At the origin, $x=y=0$, where the center
of the soliton will be placed, the lattice potential has a local minimum. It
is easy to check that the shape of the lattice different from quadratic,
adopted here (e.g., hexagonal, radial, or anisotropic) will not essentially
affect the results, cf. Ref. \cite{BBB-hex}.

In our analysis, we numerically sought for stationary localized solutions of
Eqs.~(\ref{model}), in the form of
\end{subequations}
\begin{equation}
\phi \left( x,y,t\right) =e^{-i\lambda t}u\left( x,y\right) ,\psi \left(
x,y,t\right) =e^{-i\mu t}v\left( x,y\right) ,  \label{lamu}
\end{equation}%
with corresponding chemical potentials $\lambda $ and $\mu $ and real-valued
functions $u$ and $v$ obeying the stationary equations,%
\begin{equation}
\lambda u=-\nabla ^{2}u+u^{3}-gv^{2}u-Vu,\,\ \mu v=-\nabla
^{2}v+v^{3}-gu^{2}v-Vv.\,  \label{stat}
\end{equation}

Numerical solutions were constructed on a finite-size grid that was
sufficiently large to avoid influence of boundaries on the localized
solutions. Stationary solutions were found starting from a localized input,
using the imaginary-time propagation method \cite{ITM}. The stability of these
stationary solutions is then tested by simulations of perturbed evolution of
Eqs.~(\ref{model}) in real time.

\subsection*{Symmetric solitons and vortices}

First, we look for symmetric modes with identical stationary wave functions $%
u$ and $v$ of the two components, with equal norms
\begin{equation}
M=\int \int u^{2}\left( x,y\right) dxdy,~N=\int \int v^{2}\left( x,y\right)
dxdy.  \label{MN}
\end{equation}%
In this symmetric case, the overall nonlinearity present in Eqs.~(\ref{model}%
) can be characterized by a single effective coefficient $g-1$.
Thus, $g>1$
makes the effective nonlinearity self-attractive. For $g<1$ the strength of the
repulsive inter-component interaction is larger than the attractive
cross-component interaction and  thus the net effect in this case is an overall
defocusing nonlinearity. Note that the net nonlinearity vanishes in the
symmetric case for $g=1$.

Numerical results for the symmetric solutions are summarized in Fig.~\ref%
{symmetric}. We have found that the symmetric fundamental solitons [see a
typical example in Fig. \ref{symmetric}(a)] are stable in their \emph{entire
existence region}, as shown by Fig. \ref{symmetric}(b). This conclusion
complies with the fact that, for $g>1$ and $g<1$, the $N(\mu )$ dependences
obey, severally, the Vakhitov-Kolokolov (VK) and anti-VK criteria, i.e., $%
dN/d\mu <0$ and $dN/d\mu >0$, which are necessary, although not sufficient,
stability conditions for solitons supported, respectively, by attractive
\cite{VK0,VK} and repulsive \cite{anti-VK} nonlinearities.

In fact, the fundamental solitons found at $g<1$, which correspond to the
effective repulsive nonlinearity, are gap solitons \cite{GS}. It can be
readily checked that values of $\lambda \equiv \mu $ in Fig. \ref{symmetric}%
(b), corresponding to $g<1$, fall into the first finite bandgap of the
spectrum of the linearized symmetric Eqs.~(\ref{stat}), while those
corresponding to $g>1$ belong to the semi-infinite gap. These gaps are
separated by the (first) narrow band, which is displayed in Fig. 1(b) by a
gray-shaded stripe. Unlike the fundamental solitons, the vortices
corresponding to $g>1$ and $g<1$ fall, respectively, into the first and
second finite bandgaps, which are separated by the narrow second band, as
shown in Fig. \ref{symmetric}(g). It is usually assumed that gap solitons
have a loosely bound shape, with many inner oscillations \cite{GS}.
Nevertheless, they may also feature a tightly bound shape, similar to that
of regular solitons, and in that case they can be effectively fitted by
means of the VA \cite{gap-vortex,SKA}, similar to what occurs in the present
system.

Also included in Figs.~\ref{symmetric}(b) and \ref{symmetric}(c) are the
analytical results produced by the VA for the fundamental solitons, as
detailed in the Methods Section below. The VA-predicted results agree
extremely well with their numerical counterparts (very small deviations are
observed for stronger nonlinearity and/or larger norms).

While, as mentioned above, lattice potentials often create vortex states
with a complex structure, composed of four density peaks for the sates with
vorticity $S=\pm 1$ \cite{BBB}, simple \textit{crater-shaped} stable
vortices may be found too \cite{crater}. Symmetric crater-shaped vortices
have been found in the present case too, with a typical example displayed in
Figs. \ref{symmetric}(e,f). As Fig. \ref{symmetric}(g) shows, the version of
the VA developed for the vortex solitons (see Methods Section for details)
also provides a very good accuracy, in comparison with the numerical
findings.

As said above, all vortices (here, we consider only those with $S=\pm 1$)
have been found in the first and second finite bandgaps (but not in the
semi-infinite gap). The vortices are unstable for $g<1$, when they belong to
the second bandgap. On the other hand, Fig.~\ref{symmetric}(g) demonstrates
that families of the symmetric vortex solitons are stable at $g>1$ in the
first gap, in a finite interval of
\begin{equation}
N_{\min }(g)<N<N_{\max }(g),  \label{NNN}
\end{equation}%
where $N_{\max }$ decreases with the increase of $g$, see Fig. \ref%
{symmetric}(h). Direct simulations demonstrate that unstable vortices
spontaneously evolve into robust randomly vibrating single- or multi-peak
patterns, which are asymmetric with respect to the two components.

\subsection*{Asymmetric solitons and vortices}

\begin{figure}[t]
\centerline{\includegraphics[width=8cm]{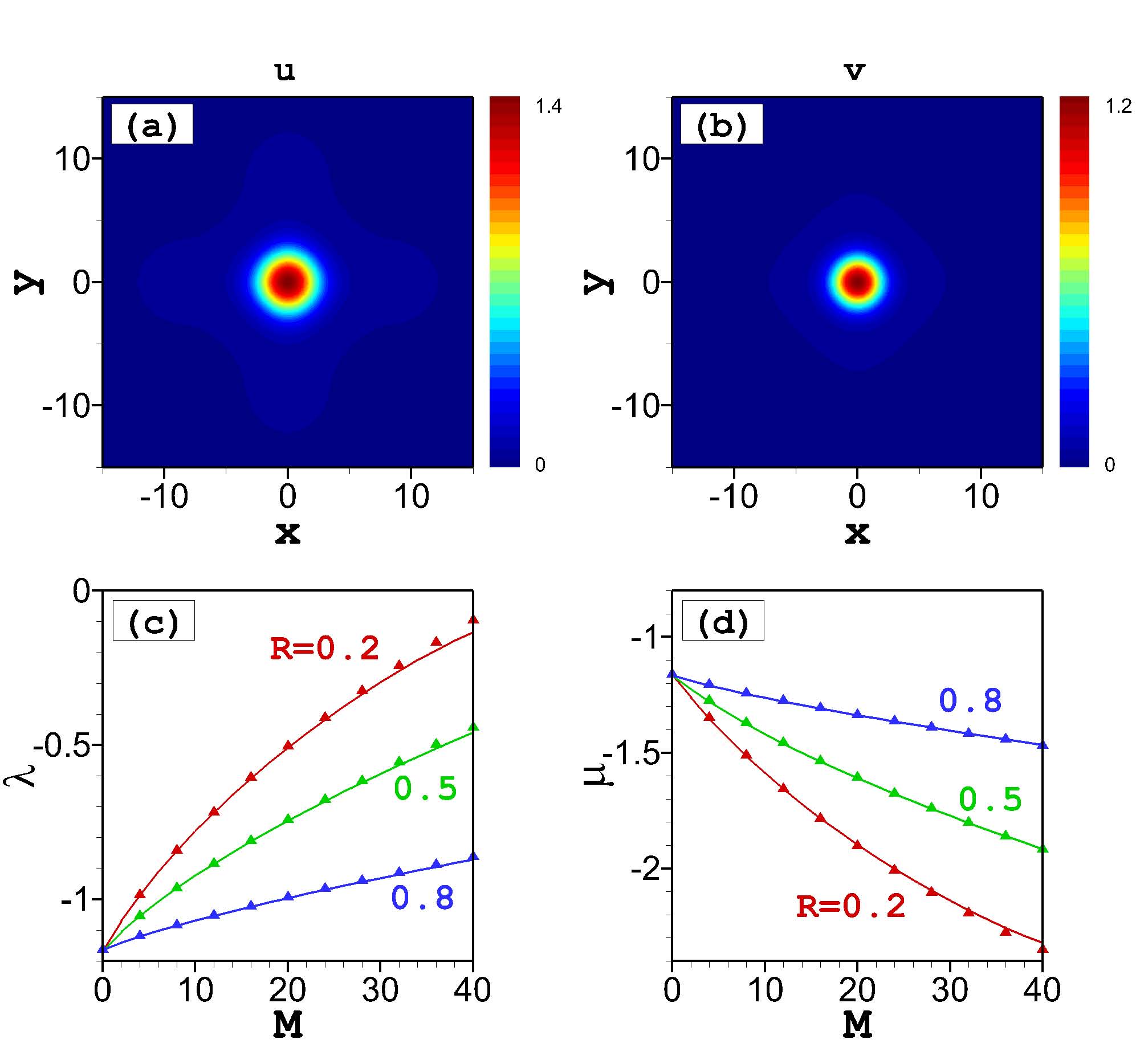}}
\caption{\textbf{Asymmetric solitons.} Amplitude distributions of the $u$-
(a) and $v$- (b) components of the stable asymmetric fundamental soliton
with $g=1$ and $N=10,$ $M=20$ (i.e., the asymmetry ratio is $R=0.5$). Panels
(c) and (d) display dependences between the larger norm ($M$) and the two
chemical potentials, $\protect\lambda $ and $\protect\mu $, respectively
[see Eqs. (\protect\ref{MN}) and (\protect\ref{lm})], for different values
of the asymmetric ratio, $R$, at $g=1$. Solid lines represent stable
numerical solutions, and triangles depict the VA-predicted analytical
results from Eqs. (\protect\ref{ab}) and (\protect\ref{lm}).}
\label{asymmetricFun}
\end{figure}

Asymmetric solitons, with unequal $u$ and $v$ components, form an especially
interesting class of modes in the present model. We start their consideration
with the case of $g=1$, in which symmetric solitons cannot exist. Basic
results for this case are displayed in Fig.~\ref{asymmetricFun}. Note that
the width of the profile with the larger norm [Fig. \ref{asymmetricFun}(a)]
is larger than that of its counterpart [Fig. \ref{asymmetricFun}(b)] with
smaller norm. For a quantitative analysis we define the asymmetry ratio, $%
R\equiv N/M$. Then, with $g=1$ and the asymmetry ratio $0<R<1$, i.e.,
roughly speaking, $|\psi |^{2}>|\phi |^{2}$, Eqs. (\ref{model:A}) and (\ref%
{model:B}) suggest that the defocusing and focusing nonlinearities dominate
in the former and latter equations, respectively. This conclusion is also
confirmed by the behavior of the chemical potentials $\lambda $ and $\mu $,
cf. their behavior in Figs. \ref{asymmetricFun} with that observed in Fig. %
\ref{symmetric}(b) for $g<1$ and $g>1$, respectively. Thus, asymmetric
solitons with $g=1$ may be considered as bound states of gap solitons and
regular ones, i.e., complexes of the \textit{semi-gap} type \cite{semigap},
the respective chemical potentials, $\lambda $ and $\mu $, falling,
respectively, into the first finite gap and semi-infinite one.

The analytical results the VA produces for these asymmetric modes are also
found to be in excellent agreement with the numerical findings, as can be
seen from the comparison of the analytical and numerical results for varying
chemical potentials $\lambda $ and $\mu $ in Figs. \ref{asymmetricFun}(c,d).
A detailed comparison (not shown here in detail) demonstrates that, as the $%
u $ component becomes broader with decrease of the asymmetry ratio $R$, this
component is more strongly affected by the underlying lattice potential,
which slightly worsens the VA accuracy for stronger nonlinearities in the
limit of $R\rightarrow 0$.

Similar to their symmetric counterparts, it has been found that the
asymmetric fundamental solitons are stable in their \emph{entire existence
region}. Intuitively, this conclusion agrees with the fact that the $\lambda
(M)$ and $\mu (M)$ dependencies in Figs. \ref{asymmetricFun}(c,d) satisfy
the anti-VK and VK criteria, respectively.

\begin{figure}[t]
\centerline{\includegraphics[width=11cm]{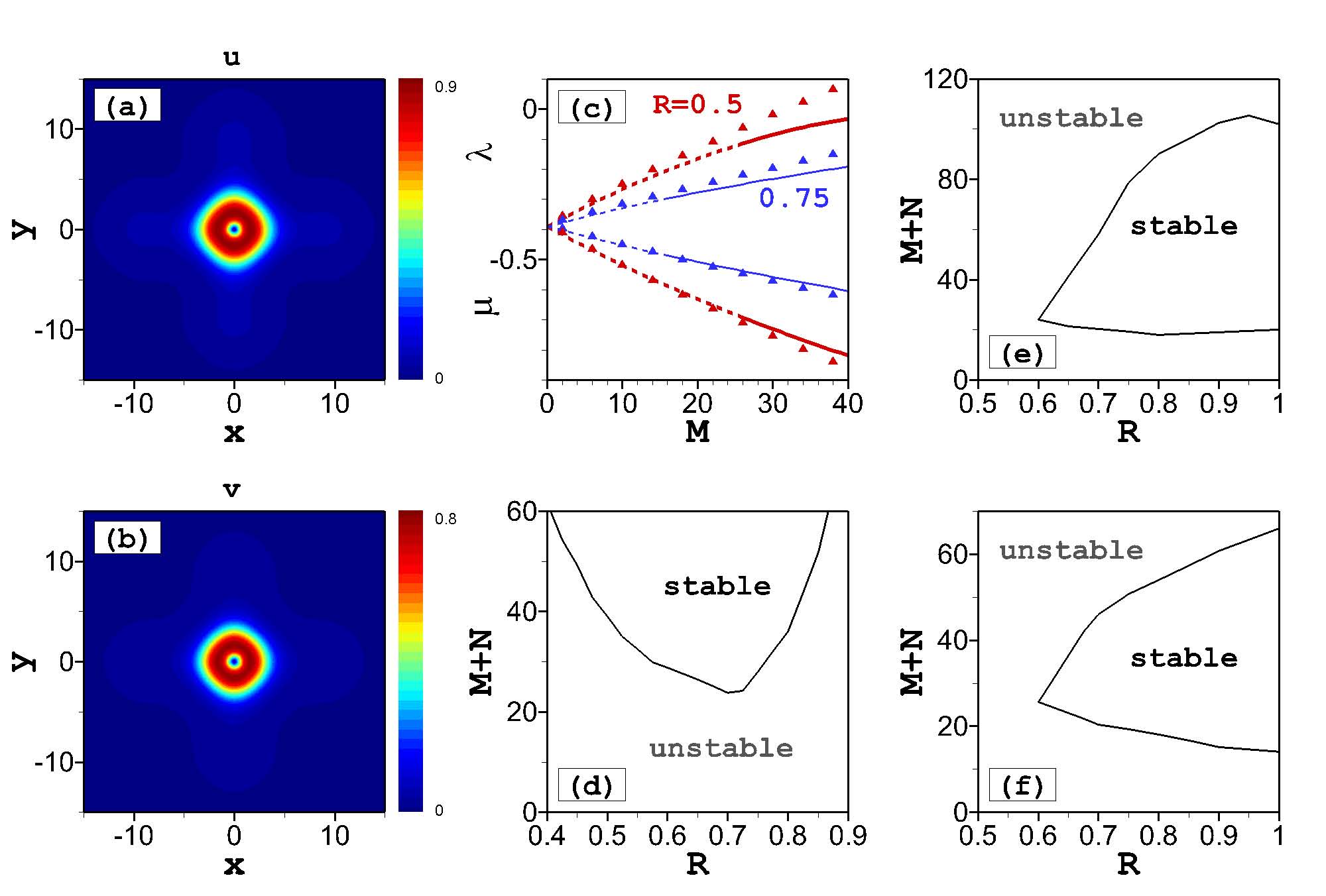}}
\caption{\textbf{Asymmetric vortices.} Amplitude distributions of the $u$-
(a) and $v$- (b) components of the asymmetric ($R=0.5$) vortex soliton at $%
g=1 $ and $M=20$. (c) Dependence of the norm ($M$) on the chemical
potentials ($\protect\mu $ and $\protect\lambda $) of the two components of
the vortex for different asymmetry ratios at $g=1$. Solid and dashed lines
denote stable and instable solutions, respectively, while triangles
represent the VA predictions, as per Eqs. (\protect\ref{abV}) and (\protect
\ref{lmV}). Regions of stability and instability for the vortices are shown
for $g=1$ in (d), $g=1.1$ in (e), and $g=1.2$ in (f).}
\label{asymmetricVor}
\end{figure}

For asymmetric vortices, the spatial profiles of the $u$ and $v$ components
feature small differences, as seen in Figs.~\ref{asymmetricVor}(a) and \ref%
{asymmetricVor}(b), while the phase distributions are virtually identical in
both components, provided that they carry the same vorticity, $S=+1$ or $-1$%
, in both components (not shown here). Similar to what is shown above for
the fundamental solitons in the case of $g=1$, when the solutions are
asymmetric ($M\neq N$), the repulsive and attractive interactions dominate
in the different components, resulting in different dependence between the
chemical potentials and the norm, as shown in Fig.~\ref{asymmetricVor}(c).

The comparison of the VA for asymmetric vortices and numerical results is
shown in Fig. \ref{asymmetricVor}(c). In the asymmetric case, the vortex
component ($u$) with a larger norm is broader than its counterpart with a
smaller norm ($v$). Also in this case, due to the influence of the
underlying lattice, the agreement of the exact numerical solution with the
simplest vortex ansatz, given by Eq. (\ref{ansatzV}), gets worse with
increasing vortex size. Therefore, the agreement of the VA and numerical
results for the $v$ component is better than for the $u$ component, see Fig. %
\ref{asymmetricVor}(c). Generally, the size of the vortices is larger than
that of the fundamental solitons, therefore the overall agreement of the
variational and numerical results for asymmetric vortices in Fig.~\ref%
{asymmetricVor}(c) is somewhat poorer than for the asymmetric fundamental
solitons in Figs.~\ref{asymmetricFun}(c) and \ref{asymmetricFun}(d).

Figure \ref{asymmetricVor}(d) shows stability and instability regions for
the vortices, depending on the asymmetry ratio, for $g=1$. At $R\rightarrow
1 $, the vortices become unstable, as the nonlinearity effectively
disappears when the solutions approach the symmetric case with $g=1$. At $%
R\rightarrow 0 $, the model reduces to a single-component system with a
defocusing nonlinearity. The smallest norm for stable vortices is found
around $R=0.7$.

Figure~\ref{asymmetricVor}(e) shows an example of the stability region for
vortices for $g>1$, \textit{viz}., with $g=1.1$. Starting from the symmetric
case with $R=1$, the stability range initially slightly widens with the
increase of the asymmetry (decrease of $R$). Then, the stability region
gradually decreases with further decrease of $R$, until all stable solutions
disappear at $R<0.6$. For $g=1.2$, as shown in Fig.~\ref{asymmetricVor}(f),
the stability region decreases monotonously with increasing asymmetry
(decrease of $R$), again shrinking to nil at $R\approx 0.6$. At $g>1$,
smaller asymmetry ratio $R$ implies domination of the attractive
nonlinearity in Eq.~(\ref{model:B}), eventually leading to an instability of
the $v$ component. It is worth noting that the fact that vortices are
unstable for larger norms in Figs.~\ref{asymmetricVor}(e) and \ref%
{asymmetricVor}(f) is in contrast to the results for $g=1$, shown in Fig.~%
\ref{asymmetricVor}(d). Roughly speaking, the stability regions for $g>1$,
shown in Figs. \ref{asymmetricVor}(e,f) are rotated by $90^{\mathrm{o}}$ in
comparison with their counterpart shown in Fig. \ref{asymmetricVor}(d) for $%
g=1$. At $g<1$, all asymmetric vortices are unstable, similar to what is
stated above for the symmetric ones.

\subsection*{Hidden-vorticity (HV) modes}

\begin{figure}[t]
\centerline{\includegraphics[width=8cm]{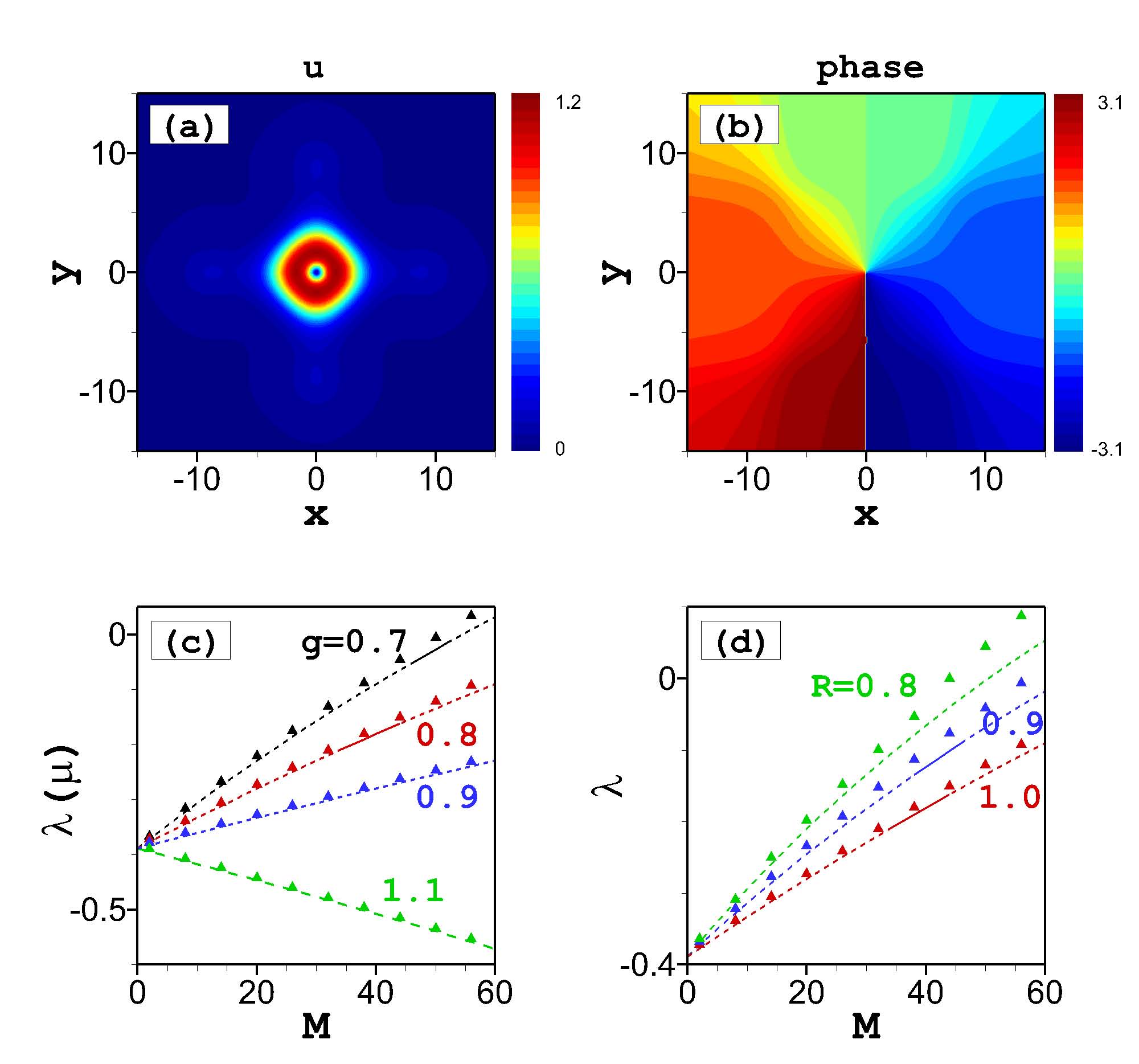}}
\caption{\textbf{Hidden vorticity modes.} (a) Amplitude and (b) phase
distributions of symmetric ($R=1$) stationary solutions for the $u$-component of
an vortex-antivortex pair at $g=0.8$ and $M=35$. Dependences of the norm ($M$%
) on the chemical potential ($\protect\lambda$) of vortex-antivortex pairs
for (c) different nonlinearity at $R=1$ and (d) different asymmetric ratio
at $g=0.8$. Solid lines are stable numerical solutions, while dashed lines
are unstable solutions. Triangles depict analytical results from Eqs. (%
\protect\ref{abV}) and (\protect\ref{lmV}).}
\label{VorAnti}
\end{figure}

For the HV vortex states, in which the two components have opposite
vorticities, results are summarized in Fig.~\ref{VorAnti}. The vorticities
of the two components are $S=+1$ and $-1$, corresponding to the phase maps
shown in Figs.~\ref{VorAnti}(b) and \ref{symmetric}(f), respectively. In the
symmetric case ($M=N$), both components have the same spatial profiles, as
shown in Fig.~\ref{VorAnti}(a). In this case, the mode carries zero angular
momentum, therefore it is called the HV state \cite{Brtka-2010-pra}.

We have found that the vortex-antivortex pairs are unstable for the
overall-focusing nonlinear system, with $g>1$, but a remarkable fact is that
they may be \emph{stable} for the defocusing system, with $g<1$, as shown in
Fig.~\ref{VorAnti}(c). Recall that all states with explicit vorticity, i.e.,
$S=1$ in both components, are entirely unstable at $g<1$, as shown above. We
have also found finite stability ranges for the vortex-antivortex pairs at $%
g<1$ in the weakly asymmetric case, with asymmetry ratios $R$ slightly
smaller than $1$, as shown in Fig.~\ref{VorAnti}(d). If the norm difference
between the two components becomes too large, the vortex-antivortex pairs
become unstable.

The VA results for the vortex-antivortex pairs are given by the same
equation, Eq.~(\ref{lmV}), as for their vortex-vortex counterparts. The
analytical results agree very well with the numerical finding when $M=N$,
see Fig. \ref{VorAnti}(c). For $M\neq N$, the difference between the
analytical and numerical results becomes more apparent due to the larger
size of the vortex components, as seen in Fig.~\ref{VorAnti}(d).

\section*{Discussion}

We have numerically and theoretically investigated the existence and
stability of symbiotic solitons and vortices in the 2D two-component spinor
system with repulsive intra-species and attractive cross-component
interactions, in the presence of an underlying lattice potential. We have
found that the fundamental solitons are \emph{always stable}, in both
symmetric and asymmetric cases (equal or different norms of the two
components). Vortex solitons with the same sign of the vorticity in both
components are stable when the attractive cross-component interaction is
stronger than the intra-species repulsion. On the other hand, the modes with
hidden vorticity, i.e., opposite signs of the vorticity in the two
components, have their stability region when the repulsive interactions
dominate. These novel types of self-trapped modes are of interest for the
understanding of general principles of pattern formation in 2D nonlinear
settings, and they can be realized in physical systems. In atomic gases, the
inter-particle interaction can be finely tuned in the vicinity of a FR
(Feshbach resonance) \cite{Chin-2010-RMP}, using the strongly dispersive
nature of the scattering length near the two-particle resonance (a negative
scattering length is associated with an attractive interaction).

This FR control may also be realized for microcavity polaritons, where the
frequency dependence of the scattering matrix element has been analyzed in
detail for excitations which are placed spectrally below the fundamental
exciton resonance \cite{Kwong-2001-PRL,Schumacher-2007-prb}. In a narrow
spectral range close to the two-particle resonance associated with the
formation of a bound biexciton state, the role of the attractive
cross-component interaction can be finely tuned so that it may dominate over
the repulsive inter-component interaction. We note that in this spectral
range, in addition to the intrinsic loss rate for polaritons, the system
will also suffer excitation-induced losses (a similar side effect of the FR
is known in atomic BECs). However, these losses may be compensated in the
presence of an exciton reservoir, or by an external pump laser, which makes
it possible to maintain the quasi-stationary behavior. In this case, the
stationary modes predicted in the present work should be observable in
semiconductor microcavities.

There remain a number of points for extension of the work, such as hybrid
bound states, with a fundamental self-trapped state in one component, and a
vortex in the other. A challenging problem is the consideration of a
three-dimensional version of the present system, which may be realized in an
atomic BEC.

\section*{Methods}


\textbf{The variational approximation (VA).} The Lagrangian of the
stationary equations (\ref{stat}) is
\begin{gather}
2L=\int \int dxdy\left\{ (\nabla u)^{2}+(\nabla v)^{2}+\frac{1}{2}\left(
u^{4}+v^{4}\right) -gu^{2}v^{2}\right.  \notag \\
\left. -\left[ \cos \left( \frac{2\pi x}{P}\right) +\cos \left( \frac{2\pi y%
}{P}\right) \right] \left( u^{2}+v^{2}\right) -\left( \lambda u^{2}+\mu
v^{2}\right) \right\} \,.  \label{L}
\end{gather}%
For fundamental solitons, the simplest Gaussian ansatz is adopted, with
amplitudes $A$, $B$ and radial widths $a$, $b$:%
\begin{equation}
u=A\exp \left( -\frac{x^{2}+y^{2}}{2a^{2}}\right) \,,~v=B\exp \left( -\frac{%
x^{2}+y^{2}}{2b^{2}}\right) \,.  \label{ansatz}
\end{equation}%
This ansatz produces the following norms, Eq.~(\ref{MN}), in the two
components: $M=\pi A^{2}a^{2}$, and $N=\pi B^{2}b^{2}$. In the asymmetric
case, $M\neq N$, we define the asymmetry ratio $R=N/M$ as noted above,
restricting it to $0<R\leq 1$. The substitution of ansatz (\ref{ansatz})
into Lagrangian~(\ref{L}) and subsequent integration yields the following
effective Lagrangian:%
\begin{gather}
2L_{\mathrm{eff}}=-\lambda M-\mu N-2M\exp \left( -\frac{\pi ^{2}a^{2}}{P^{2}}%
\right) -2N\exp \left( -\frac{\pi ^{2}b^{2}}{P^{2}}\right)  \notag \\
-\frac{g}{\pi }\frac{MN}{a^{2}+b^{2}}+\frac{M^{2}}{4\pi a^{2}}+\frac{N^{2}}{%
4\pi b^{2}}+\frac{M}{a^{2}}+\frac{N}{b^{2}}.  \label{L_eff}
\end{gather}%
The first pair of the variational equation following from Eq. (\ref{L_eff})
makes it possible to express $a^{2}$ and $b^{2}$ in terms of $M$ and $N$: $%
\partial L_{\mathrm{eff}}/\partial \left( a^{2}\right) =\partial L_{\mathrm{%
eff}}/\partial \left( b^{2}\right) =0$, i.e.,
\begin{subequations}
\label{ab}
\begin{align}
\frac{2\pi ^{2}}{P^{2}}\exp \left( -\frac{\pi ^{2}a^{2}}{P^{2}}\right) +%
\frac{g}{\pi }\frac{N}{\left( a^{2}+b^{2}\right) ^{2}}& =\frac{M+4\pi }{4\pi
a^{4}},  \label{ab:A} \\
\frac{2\pi ^{2}}{P^{2}}\exp \left( -\frac{\pi ^{2}b^{2}}{P^{2}}\right) +%
\frac{g}{\pi }\frac{M}{\left( a^{2}+b^{2}\right) ^{2}}& =\frac{N+4\pi }{4\pi
b^{4}}.  \label{ab:B}
\end{align}%
For given $M$ and $N$, $a^{2}$ and $b^{2}$ can be found by means of a
numerical solutions of the algebraic system (\ref{ab}). The chemical
potentials, $\lambda $ and $\mu $, do not appear in Eq. (\ref{ab}). They are
produced by the second pair of the variational equations: $\partial L_{%
\mathrm{eff}}/\partial M=\partial L_{\mathrm{eff}}/\partial N=0$, i.e.,
\end{subequations}
\begin{subequations}
\label{lm}
\begin{align}
\lambda & =-2\exp \left( -\frac{\pi ^{2}a^{2}}{P^{2}}\right) -\frac{g}{\pi }%
\frac{N}{a^{2}+b^{2}}+\frac{M+2\pi }{2\pi a^{2}},  \label{lm:A} \\
\mu & =-2\exp \left( -\frac{\pi ^{2}b^{2}}{P^{2}}\right) -\frac{g}{\pi }%
\frac{M}{a^{2}+b^{2}}+\frac{N+2\pi }{2\pi b^{2}}.  \label{lm:B}
\end{align}%
The first objective of the VA is, for given $g$, to identify a region in the
plane of $\left( M,N\right) $ where the numerical solution of Eq. (\ref{ab})
produces physically relevant solutions, with $a^{2},b^{2}>0$.

Similarly to the approach for the fundamental solitons outlined above, for
vortices we adopt a natural generalization of the Gaussian ansatz:
\end{subequations}
\begin{equation}
u=A(x+iy)\exp \left( -\frac{x^{2}+y^{2}}{2a^{2}}\right) ,~v=B(x+iy)\exp
\left( -\frac{x^{2}+y^{2}}{2b^{2}}\right) ,  \label{ansatzV}
\end{equation}%
whose norms are $M=\pi A^{2}a^{4}$ and $N=\pi B^{2}b^{4}$. Substituting
ansatz (\ref{ansatzV}) into Lagrangian (\ref{L}), the following effective
Lagrangian is obtained, cf. Eq. (\ref{L_eff}):%
\begin{gather}
2L_{\mathrm{eff}}=-\lambda M-\mu N-\frac{g}{\pi }\frac{2a^{2}b^{2}MN}{%
(a^{2}+b^{2})^{3}}+\frac{M^{2}}{8\pi a^{2}}+\frac{N^{2}}{8\pi b^{2}}+\frac{2M%
}{a^{2}}+\frac{2N}{b^{2}}  \notag \\
-\frac{2M}{P^{2}}\left( P^{2}-a^{2}\pi ^{2}\right) \exp \left( -\frac{\pi
^{2}a^{2}}{P^{2}}\right) -\frac{2N}{P^{2}}\left( P^{2}-b^{2}\pi ^{2}\right)
\exp \left( -\frac{\pi ^{2}b^{2}}{P^{2}}\right) .
\end{gather}%
Then, the variational equations for $a^{2}$ and $b^{2}$ are produced as $%
\partial L_{\mathrm{eff}}/\partial \left( a^{2}\right) =\partial L_{\mathrm{%
eff}}/\partial \left( b^{2}\right) =0$, i.e.,
\begin{subequations}
\label{abV}
\begin{align}
\frac{2\pi ^{2}\left( 2P^{2}-a^{2}\pi ^{2}\right) }{P^{4}}\exp \left( -\frac{%
\pi ^{2}a^{2}}{P^{2}}\right) +\frac{g}{\pi }\frac{2b^{2}N\left(
2a^{2}-b^{2}\right) }{\left( a^{2}+b^{2}\right) ^{4}}& =\frac{M+16\pi }{8\pi
a^{4}},  \label{abV:A} \\
\frac{2\pi ^{2}\left( 2P^{2}-b^{2}\pi ^{2}\right) }{P^{4}}\exp \left( -\frac{%
\pi ^{2}b^{2}}{P^{2}}\right) +\frac{g}{\pi }\frac{2a^{2}M\left(
2b^{2}-a^{2}\right) }{\left( a^{2}+b^{2}\right) ^{4}}& =\frac{N+16\pi }{8\pi
b^{4}}.  \label{abV:B}
\end{align}%
Finally, the chemical potentials for the two components of the vortex are
produced according by the remaining variational equations, $\partial L_{%
\mathrm{eff}}/\partial M=\partial L_{\mathrm{eff}}/\partial N=0$, i.e.,
\end{subequations}
\begin{subequations}
\label{lmV}
\begin{align}
\lambda & =\frac{2\left( a^{2}\pi ^{2}-P^{2}\right) }{P^{2}}\exp \left( -%
\frac{\pi ^{2}a^{2}}{P^{2}}\right) -\frac{g}{\pi }\frac{2a^{2}b^{2}N}{\left(
a^{2}+b^{2}\right) ^{4}}+\frac{M+8\pi }{4\pi a^{2}},  \label{lmV:A} \\
\mu & =\frac{2\left( b^{2}\pi ^{2}-P^{2}\right) }{P^{2}}\exp \left( -\frac{%
\pi ^{2}b^{2}}{P^{2}}\right) -\frac{g}{\pi }\frac{2a^{2}b^{2}M}{\left(
a^{2}+b^{2}\right) ^{4}}+\frac{N+8\pi }{4\pi b^{2}}.  \label{lmV:B}
\end{align}


\begin{center}
\textbf{Acknowledgements}
\end{center}

{The Paderborn groups acknowledge support from the Deutsche
Forschungsgemeinschaft (DFG) through the collaborative research center TRR~142. S.S.
further acknowledges support through the Heisenberg programme of the DFG.}
B.A.M. appreciates hospitality of the Department of Physics and the Center for
Optoelectronics and Photonics Paderborn (CeOPP) at the University of
Paderborn.

\begin{center}
\textbf{Author contribution}
\end{center}

X.M. and R.D. performed the simulations. B.A.M. and X.M. derived the
analytical results. X.M., B.A.M., and S.S. prepared the manuscript. S.S. and
B.A.M. conceived the idea. All authors analyzed the data and reviewed the
manuscript. 


\begin{center}
\textbf{Additional information}
\end{center}

Competing financial interests: The authors declare no competing financial
interests. 
\end{subequations}

\end{document}